\documentclass[smallextended]{svjour3}       % onecolumn (second format)
\usepackage{fullpage, amsmath, amssymb, array, bbm}
\usepackage{natbib,enumerate}

\smartqed  % flush right qed marks, e.g. at end of proof
\usepackage{graphicx}
 \journalname{Journal of Computational Neuroscience}
\begin{document}

\title{Firing Rate Dynamics in Recurrent Spiking Neural Networks with Intrinsic and Network Heterogeneity}


\titlerunning{Intrinsic and Network Heterogeneity}        % if too long for running head

\author{Cheng Ly
}


\institute{C. Ly \at
	     Department of Statistical Sciences and Operations Research \\
	     Virginia Commonwealth University \\
              Richmond, Virginia 23284-3083 USA \\
              Tel.: +1(804) 828-5842\\
              Fax: +(804) 828-8785\\
              \email{CLy@vcu.edu}           %  \\
}

\date{Received: date / Accepted: date}

\maketitle

%250 word limit
\begin{abstract}
Heterogeneity of neural attributes has recently gained a lot of attention and is increasing recognized as a crucial feature in neural processing. 
Despite its importance, this physiological feature has traditionally been neglected in theoretical studies of cortical neural networks. 
Thus, there is still a lot unknown about the consequences of cellular and circuit heterogeneity in spiking neural networks. 
In particular, combining network or synaptic heterogeneity and intrinsic heterogeneity has yet to be considered systematically despite the fact that both are known to exist and 
likely have significant roles in neural network dynamics. In a canonical recurrent spiking neural network model, we study how these two forms of heterogeneity 
lead to different distributions of excitatory firing rates. To analytically characterize how these types of heterogeneities affect the network, we employ a dimension reduction method that relies 
on a combination of Monte Carlo simulations and probability density function equations. We find that the relationship between intrinsic and network heterogeneity has a strong effect on 
the overall level of heterogeneity of the firing rates. Specifically, this relationship can lead to amplification or attenuation of firing rate heterogeneity, and 
these effects depend on whether the recurrent network is firing asynchronously or rhythmically firing. These observations are captured with the aforementioned reduction method, and furthermore 
simpler analytic descriptions based on this dimension reduction method are developed. The final analytic descriptions provide compact and descriptive formulas for how the relationship between 
intrinsic and network heterogeneity determines the firing rate heterogeneity dynamics in various settings.

\end{abstract}

\section{Introduction}
\label{intro}

Theoretical studies of spiking neuronal networks have been extremely valuable for experimentalists and theoreticians.  Uncovering the underlying mechanisms of complex phenomena 
in neural circuits often requires theory and/or computation.  In this vein, this paper focusses on the effects of heterogeneous neural attributes in model neural networks.  
%There are a wide variety of phenomena that can be well described theoretically without addressing some of the messy biological details \citep{laing_lord09,bardBook,prcBook}.  
Heterogeneity is an undeniable physiological feature that has often been ignored in theoretical studies because it complicates theoretical analyses.  Not only is there ample experimental 
evidence for heterogeneity of neural networks at many scales \citep{markram1997,parker2003variable,marder06,bremaud_07}, but the importance of heterogeneity in 
neural computational is becoming more apparent.  Indeed, a combination of theoretical 
and experimental studies on neural networks have demonstrated the value of heterogeneity \citep{ostojicBH09,hermann12}.  In particular, theoretical studies have shown that heterogeneity can 
generally lead to efficient neural coding \citep{shamir06,chelaru08,padmanabhan10,marsat10,mejias12,tripathy13}.  However, there is still not a lot known about heterogeneity.  Specifically, how do 
different sources of heterogeneity interact and lead to different neural network activity?

We make a distinction between different sources of heterogeneity, addressing two forms: intrinsic and network heterogeneity, both of which are known to exist.  Intrinsic heterogeneity 
are differences due to cellular properties that exist without coupling to other neurons \citep{marder06,padmanabhan10}, 
for example the membrane time constant, threshold for spiking, reversal potentials, etc.  
Network heterogeneity, that is heterogeneity induced by coupling in a neural network, can arise from differences in synaptic coupling between neurons \citep{parker2003variable,marder06,bremaud_07,oswald09}.  
To the best of our knowledge, the physiological relationship between these two sources of heterogeneity are not known.  Therefore, we systematically study the effects of these forms of 
heterogeneity on a canonical recurrent spiking neural network.  The main motivation for studying the relationship between different heterogeneous components is to provide a framework for possibly 
reconciling experimental measurements of multiple neural attributes; recent theoretical studies have shown that the network components can interact nonlinearly with surprising results 
\citep{marder06,mejias14,hunsberger14} (see Discussion section).  
The results of our study clearly show how multiple components effect the firing rate variability and might apply to experiments that measure the heterogeneity of these 
(or possibly other) neural attributes.

The network we consider is noisy with variable spiking similar to that of real cortical neurons, and is excitable (i.e., neurons only fire with noise and/or synaptic coupling).  
We analyze how intrinsic and network heterogeneity together alter the dynamics of strongly coupled networks of neurons in various regimes ranging from asynchronous to 
rhythmic (i.e, 'ping' network) firing.  We focus on the dynamics of the excitatory neurons because they are the predominate cells for propagating signals to different layers in the cortex.  
Unsurprisingly, we find that when both intrinsic and network heterogeneity increase independently (i.e., when there is no relationship between them), the excitatory firing 
rates tend to have a larger range.  However, for a fixed level of heterogeneity, the relationship or correlation between intrinsic and network heterogeneity strongly affects the overall range of excitatory firing 
rates.  Moreover, these effects depend on what regime the neural network is in: during rhythmic firing, excitatory firing rate ranges decrease when intrinsic and network heterogeneity correlation increases; during asynchronous firing, excitatory firing rate ranges increase when intrinsic and network heterogeneity correlation increases.

To better understand these observations, we implement a dimension reduction method that relies on a combination of Monte Carlo simulations and analytic reductions.  
The reduction theory is based in part on our previous work \citep{Ly_PrincDimRed_13,nlc_15}, and on the work of others \citep{moreno3,nesse08}, where particular state variables are assumed to be slow and thus 
decoupled from other variables.  Fortunately, the dimension reduction method captures the qualitative trend of the range of excitatory firing rates as heterogeneity is changed.  This further 
inspires a simpler yet more revealing analytic description for how intrinsic and network heterogeneity combine to yield different ranges of excitatory firing rates.  This study gives a more complete understanding of 
how heterogeneities interact and result in modulation of the firing rate statistics, which may ultimately lead to a better understanding of neural coding in coupled neural networks.

%\citep{markram1997,oswald09,xue14}

\section{Methods}

\subsection{Recurrent spiking LIF network}
\label{sec:lif}

The recurrent spiking network of excitatory (E) and inhibitory (I) neurons are modeled as leaky integrate-and-fire (LIF) neurons.  
The intrinsic and network heterogeneity are modeled simply by two parameters that are allowed to vary among the neurons.  Other 
modeling studies impose heterogeneity in the response property \citep{tripathy13}, e.g., orientation tuning \citep{shamir06,chelaru08}.  
The models here effectively have heterogeneous response properties, but our focus is on two different sources that lead to that property.  
We model the intrinsic heterogeneity by having different voltage thresholds for spiking \citep{mejias12,yim13}, equivalent to how many have incorporated intrinsic heterogeneity 
\citep{strogatzmirollo,chow98,burton12,LMD_whisker_12}.  The network heterogeneity is modeled by scaling the synaptic input by a value, effectively making each neuron receive 
different levels of network input.  There is evidence in slice recordings that the probability of connection depends on distance \citep{oswald09,levy12}, although we are not 
taking into account spatial dynamics, this model is plausible assuming synaptic strengths do not all inversely scale with connection probability.  
Moreover, there is abundant evidence for differences in synaptic coupling between neurons \citep{parker2003variable,marder06,bremaud_07}.  

The equations for the excitatory neurons indexed by $j\in\{1,2,\dots,N_e\}$ are:
\begin{eqnarray}\label{e_lif}
	\tau_m \frac{d v_j}{dt} & = & -v_j-g_{ie}(t)(v_j-\mathcal{E}_I)-g_{ee}(t)(v_j-\mathcal{E}_E)+\sigma_E\eta_j(t) \nonumber \\
	v_j(t^*) & \geq & \theta_j \hbox{(refractory period)} \Rightarrow v_j(t^*+\tau_{ref})=0 \nonumber \\
	\tau_n \frac{d \eta_j}{dt} & = & -\eta_j+\sqrt{\tau_n}\xi_j(t) \nonumber \\
	g_{ee}(t) &=& q_j \frac{\gamma_{ee}}{p_{ee}N_e}\sum_{j'\in\{\hbox{presyn E cells}\}} G_{j'}(t) \nonumber \\
	g_{ei}(t) &=& \frac{\gamma_{ei}}{p_{ei}N_i}\sum_{k'\in\{\hbox{presyn I cells}\}} G_{k'}(t) \nonumber \\
	\tau_d\frac{d G_j}{dt} &=& -G_j + A_j  \nonumber \\
	\tau_r\frac{d A_j}{dt} &=& -A_j + \tau_r \alpha \sum_{l} \delta(t-t_l)
\end{eqnarray}
where the leak, inhibitory and excitatory reversal potentials are 0, $\mathcal{E}_I$, and $\mathcal{E}_E$, respectively with $\mathcal{E}_I<0<\mathcal{E}_E$ 
(the voltage is scaled to be dimensionless so that a leak/resting value of -65\,mV maps to 0 and a threshold voltage of -55\,mV maps to 1 (see Table \ref{tab:1})).  
$\xi_j(t)$ are uncorrelated white noise processes, $p_{xy}$ is the proportion of neuron type $y$ (randomly chosen) that provides presynaptic input to neuron type $x$ ($x,y\in\{e,i\}$).  The second line 
in the equations describes the refractory period at spike time $t^*$: when the neuron's voltage crosses 
threshold $\theta_j$ ({\bf intrinsic heterogeneity}), 
the neuron goes into a refractory period for $\tau_{ref}$ where the voltage is undefined\footnote{In refractory, the other variables are governed by their ODEs}, 
after which we set the neuron's voltage to 0. In the last line, $t_l$ denotes the spike times of the $j^{th}$ excitatory neuron. There are two factors in the equation for the total synaptic 
conductances ($g_{ee}$ and $g_{ei}$): $q_j$ and $\frac{\gamma_{xy}}{p_{xy}N_y}$; the latter does not depend on the individual neuron and is the same across the entire (E) population. 
However, $q_j$ introduces {\bf network heterogeneity} by scaling {\it both} excitatory and inhibitory synaptic inputs.  This form of network heterogeneity is loosely motivated by recent results by 
\citet{xue14}, who found that pyramidal neurons receive relatively similar proportions of excitation and inhibition in layer 2/3 of mammalian visual cortex (i.e., some cells receive more E/I while some 
cells receive less E/I).  The $q_j$ factors are a straight forward way to capture the different levels of 'balanced' input (see equation \eqref{i_lif}).  The term network heterogeneity is often used to mean 
that the structure of the network is heterogeneous; here, we use that term to mean that the network activity induces heterogeneity via network inputs.

Similarly, for the inhibitory neurons indexed by $k\in\{1,2,\dots,N_i\}$, the equations are:
\begin{eqnarray}\label{i_lif}
	\tau_m \frac{d v_k}{dt} & = & -v_k-g_{ii}(t)(v_k-\mathcal{E}_I)-g_{ei}(t)(v_k-\mathcal{E}_E) +\sigma_I\eta_k(t) \nonumber \\
	v_k(t^*) & \geq & 1 \hbox{(refractory period)} \Rightarrow v_j(t^*+\tau_{ref})=0 \nonumber \\
	\tau_n \frac{d \eta_k}{dt} & = & -\eta_k+\sqrt{\tau_n}\xi_k(t) \nonumber \\
	g_{ie}(t) &=&  q_j\frac{\gamma_{ie}}{p_{ie}N_e}\sum_{k'\in\{\hbox{presyn I cells}\}} G_{k'}(t) \nonumber \\
	g_{ii}(t) &=&  \frac{\gamma_{ii}}{p_{ii}N_i}\sum_{k'\in\{\hbox{presyn I cells}\}} G_{k'}(t) \nonumber \\
	\tau_d\frac{d G_k}{dt} &=& -G_k + A_k  \nonumber \\
	\tau_r\frac{d A_k}{dt} &=& -A_k + \tau_r \alpha \sum_{l} \delta(t-t_l) 
\end{eqnarray}
Notable differences compared to excitatory neurons are that the threshold values are all equal to 1, and there is not a $q_k$ factor that scales the presynaptic inputs from the network.  
Although one could in principle relax these assumptions and augment the subsequent theory in a standard way, we made this choice because the results in this paper would not be diminished, and to avoid 
distracting from our focus on excitatory neuron behavior.  The parameter values for all of the figures are in Table \ref{tab:1}.

We consider two regimes of this model: (i) {\bf noisy rhythm}, where the power spectrum is larger for certain frequency values, and (ii) {\bf asynchronous} that has a flat power spectrum.  Figure 
\ref{fig:1} shows a comparison of various quantities in the three regimes considered in this paper (see Table \ref{tab:2} for parameter values of different regimes).  

\begin{table}
\caption{Parameters for all simulations}
\label{tab:1}       % Give a unique label
\begin{tabular}{lllllll}
\hline\noalign{\smallskip}
{\bf Parameter} & $\tau_m$ & $\tau_{ref}$ & $\mathcal{E}_I$ & $\mathcal{E}_E$ & $\tau_n$ & $p_{xy}$ \\
\noalign{\smallskip}\hline\noalign{\smallskip}
For E and I: & 20\,ms & 2\,ms & -0.5 & 6.5 & 5\,ms & 0.2 \\
%I cells & 20\,ms & 2\,ms & -0.5 & 6.5 & 5\,ms & 0.2\\
\noalign{\smallskip}\hline
\end{tabular}

\begin{tabular}{lllllll}
\hline\noalign{\smallskip}
{\bf Parameter} & $\tau_d$ & $\tau_r$ & $\alpha$ & $N_{e/i}$ \\
\noalign{\smallskip}\hline\noalign{\smallskip}
E cells & 1\,ms & 5\,ms & 1 & 800  \\
I cells & 2\,ms & 10\,ms & 2 & 200 \\
\noalign{\smallskip}\hline
\end{tabular}

\end{table}

\begin{table}
\caption{Parameters for specific regimes}
\label{tab:2}       % Give a unique label
\begin{tabular}{lllllll}
\hline\noalign{\smallskip}
{\bf Regime:} & $\gamma_{ei}$ & $\gamma_{ie}$ & $\gamma_{ee}$ & $\gamma_{ii}$ & $\sigma_E$ & $\sigma_I$\\
\noalign{\smallskip}\hline\noalign{\smallskip}
{\bf Noisy Rhythm } & 10 & 8 &  12.25 & 5  & 2.5 &  3 \\
{\bf Asynchronous} & 10 & 8 &  0.05 & 5  & 3.5 &  4 \\
{\bf Sharp Rhythm} & 10 & 8 &  11.5 & 5  & 2.55 &  -- \\
\noalign{\smallskip}\hline
\end{tabular}
\end{table}

\paragraph{Monte Carlo Simulations:}  Monte Carlo simulations were run for 100\,s of simulation time for ten realizations (Fig. \ref{fig:1} has only one realization) with a time step $\Delta t=0.2$\,ms, 
and the firing rates statistics were binned in non-overlapping 1\,ms time windows.  We use a common estimate of the standard deviation of the firing rate calculations with Monte Carlo simulations 
across the $n=10$ realizations: 
$$\overline{\sigma_{\nu(j)}}\approx \sqrt{ \frac{1}{n-1} \sum_{m=1}^n \left( \nu_m(j)-\overline{\nu_m(j)} \right)^2 };$$  
the gray shaded regions in panels A and C of Figures \ref{fig:2}--\ref{fig:4} represent 1 standard deviation above and below the sample mean: $\overline{\nu_m(j)} \pm \overline{\sigma_{\nu(j)}}$.  
The Monte Carlo simulation plots in panel B of Figures \ref{fig:2}--\ref{fig:4} did not include standard deviation regions because the plots would be harder to see.  The error bars of the Monte Carlo 
simulations in panel D of Figures \ref{fig:2}--\ref{fig:4} and Figure \ref{fig:5}A, C, E, represent an estimate of the standard deviation of: $\displaystyle\max_j{\nu}-\displaystyle\min_j{\nu}$.  We use the following estimate:
$$\overline{\sigma_{\nu(\max)-\nu(\min)}} \approx \sqrt{\overline{\sigma_{\nu(\max)} }^2 + \overline{\sigma_{\nu(\min)} }^2 } $$ 
even though $\overline{\sigma_{\nu(\max)-\nu(\min)}}=\sqrt{\overline{\sigma_{\nu(\max)} }^2 + \overline{\sigma_{\nu(\min)} }^2 -2Cov(\nu(\max),\nu(\min))}$; the reason for this is because estimating 
 $2Cov(\nu(\max),\nu(\min))$ would require storing an additional $O(N_e^2)$ sized vector, as well as $O(n^2)$ simulations for similar order accuracy (for every single parameter set), 
 and the computation times are already quite long.  

\subsection{Model with a sharper rhythm}

Another model considered is one with less variable inhibitory firing that ultimately leads to sharper rhythms in the excitatory neurons.  The reason for an observed sharper rhythm 
(Fig. \ref{fig:1}F) is that in these recurrent networks, 
the inhibitory neuron firing silence and thus shape the rhythm of the excitatory neurons \citep{borgers03,economo12}.  The only change is in the inhibitory 
neuron's voltage equation which no longer has a noise term, but rather a deterministic drift to a sub-threshold target voltage $\mathcal{E}_{det}$:
\begin{equation}
	\tau_m \frac{d v_k}{dt} = -v_k-g_{ii}(t)(v_k-\mathcal{E}_I)-g_{ie}(t)(v_k-\mathcal{E}_E) -g_{det}(v_k-\mathcal{E}_{det}) \nonumber \\
\end{equation}
we set $\mathcal{E}_{det}=0.9$ and $g_{det}=2$.  The regime considered has a strong and relatively regular oscillation (Figure \ref{fig:1} right column {\bf sharp rhythm}).

%\paragraph{Paragraph headings.} Use paragraph headings as needed.
%\section{Different levels of heterogeneity}
In addition to showing distinct characteristics compared to the other regimes (noisy rhythm, asynchronous), the main motivation for considering this additional model is that such recurrent 
networks with a sharp gamma rhythm are commonly studied and known to be important for coding in many 
areas of the cortex \citep{borgers03,wangReview10,buzsaki12,economo12}.

The following two subsections describe the way both network and intrinsic heterogeneity are modulated in this paper.

\subsection{Changing the level of intrinsic and network heterogeneity independently}

The two heterogeneous parameters $(q_j,\theta_j)$ are varied to yield significant changes in the range of firing rates.  The means of both $\vec{q}$ and $\vec{\theta}$ are set to 1, and the 
parameters $\sigma_q\in[0,1]$ and $\sigma_\theta\in[0,1]$ 
quantify the level of the network and intrinsic heterogeneities, respectively, in the following way:
\begin{eqnarray}\label{het_sig1}
	\vec{q} &\sim& 1+\sigma_q*(\mathcal{U}-0.5) \\
	\vec{\theta} &\sim& 1+\sigma_\theta*\mathcal{N} \label{het_sig2}
\end{eqnarray}
where $\mathcal{U}$ is the uniform distribution on $[0,1]$, and $\mathcal{N}$ is a truncated\footnote{The middle 98.76\% is included, so for 
$\sigma_\theta=1$, $\vec{\theta}\in[0.8,1.2]$} normal distribution with mean 0 and standard deviation 0.08.  When chosen independently, the correlation between these two vectors will 
be small and theoretically zero.

\subsection{Changing the correlation between intrinsic and network heterogeneity}\label{sec:corr_sig}

We consider another way to change the heterogeneity where the overall level is approximately the same but the correlation between $\vec{q}$ and $\vec{\theta}$ is set to 
a prescribed value.  Given the vectors $\vec{q}$ and $\vec{\theta}$, we fix $\vec{q}$ to the same values but transform $\vec{\theta}$ so that the Pearson's correlation coefficient is 
$\varrho\in(-1,1)$ in such a way that the transformed vector has the same mean and variance as $\vec{\theta}$.  The details for how this is accomplished are described in the Appendix.  

\subsection{High dimensional probability density equation}
\label{sec:hd_pdf}

The recurrent coupled stochastic network in section \ref{sec:lif} is difficult to describe theoretically.  A common method uses probability density functions (p.d.f.), or 
a population density methods, where the probability of a neuron being in a particular state has a corresponding equation.  The variables in the populations are no longer 
tracked individually, but rather captured by a p.d.f.; for example, $(V_j,G_j,A_j,\eta_j)$ for $j=1,2,\dots,N_e$ are captured with a function of $(v_E,g_E,a_E,\eta_E)$.  The 
two forms of heterogeneity introduce even more dimensions than the usual state variables.  For simplicity, one can track a family of probability density functions for each $(q_j,\theta_j)$ pair or each 
distinct neuron.  The subsequent equations are a good approximation to the coupled network \eqref{e_lif}-\eqref{i_lif} with the following assumptions: 
\begin{enumerate}[(i)]
	\item finite size effects are negligible ($N_{e/i}\gg1$)
	\item the firing rate of presynaptic neurons is governed by a Poisson process
	\item the population firing rate averaged over $\vec{q}$ and $\vec{\theta}$ is a good approximation to the average presynaptic input rate
	\item a single p.d.f. function is adequate to describe the population behavior, and the heterogeneity is driven by $(q_j,\theta_j)$
\end{enumerate}
The first two assumptions are standard in this framework, while the last two assumptions has been successfully used \citep{Ly_heterOsc_14}, 
where a family of probability density functions were indexed by the quenched heterogeneity values.  
Even though these assumptions are violated, the following equations are key for the reduced descriptions in sections \ref{sec:red_theory}, \ref{sec:thry_h1}--\ref{sec:thry_h2}.

For each pair of values $(q_j,\theta_j)$, the probability density function $\rho$ is defined by:
\begin{eqnarray}
\int_\Omega
	\rho (v_E,\vec{w}_E,v_I,\vec{w}_I,t)\,dv_E\,d\vec{w}_E\,dv_I d\vec{w}_I &=& \Pr\Big(\big(v_E(t),\vec{w}_E(t),v_I(t),\vec{w}_I(t)\big)\,\,\in\,\,\Omega \Big) \nonumber 
\end{eqnarray}
where $\vec{w}_X$ denotes the other states variables of the corresponding neuron type $X\in\{E,I\}$, 
consisting of conductance, colored noise: $\vec{w}_X=(g_X,a_X,\eta_X)$.  The evolution of the p.d.f.'s is 
governed by a continuity equation and boundary conditions:

\begin{eqnarray} \label{full_fp_eqn}
 \frac{\partial \rho}{\partial t} &=& - \nabla \cdot {\vec{J}} \\
 	\vec{J} & := & (J_{v_E},J_{g_E},J_{a_E},J_{\eta_E},J_{v_I},J_{g_I},J_{a_I},J_{\eta_I}) \\
	J_{v_E} &:=& -\frac{1}{\tau_m}\left[v_E+q\gamma_{ei} g_I (v_E-\mathcal{E}_I)+q\gamma_{ee} g_E(v_E-\mathcal{E}_E)+\sigma_E\eta_E \right] \rho \\
	J_{v_I} &:=& -\frac{1}{\tau_m}\left[v_I+\gamma_{ii} g_I (v_I-\mathcal{E}_I)+\gamma_{ie} g_E(v_I-\mathcal{E}_E)+\sigma_I\eta_I \right] \rho \\
	J_{g_X} &:=& -\frac{1}{\tau_d}\left[g_X-a_X\right] \rho \\
	J_{a_X} &:=& -\frac{a_X}{\tau_r} + \nu_X(t)\int_{a_X-\alpha_X}^{a_X} \rho(\dots,a_X',\dots)\,da_X' \\
	J_{\eta_X} &:=& -\frac{1}{\tau_n} \eta_X\rho + \frac{1}{\tau_n}\frac{\partial^2 \rho}{\partial \eta_X^2} \\
	\nu_X(t) &:=& \frac{1}{\tau_m} \int\int\int J_{v_X}(v_X=\theta,\vec{w}_X) \,d\vec{w}_X\,dq\,d\theta \label{nu_eqn} \\
	J_{v_X}(v_X=\theta,\vec{w}_X,t) &=& J_{v_X}(v_X=0,\vec{w}_X,t+\tau_{ref}) \\
	J_{\vec{w}_X}\vert_{\partial \vec{w}_X}     &=& 0 \label{last_bc}
\end{eqnarray}
%	J_{g_E} &:=& -\frac{1}{\tau_d}\left[g_E-a_E\right] \rho \\
%	J_{a_E} &:=& -\frac{a_E}{\tau_r} + \nu_E(t)\int_{a_E-\alpha_E}^{a_E} \rho(v_E,g_E,a_E',\eta_E,v_I,g_I,a_I,\eta_I,t)\,da_E' \\
%	J_{\eta_E} &:=& -\frac{1}{\tau_n} \eta_E\rho + \frac{1}{\tau_n}\frac{\partial^2 \rho}{\partial \eta_E^2} \\
%	J_{g_I} &:=& -\frac{1}{\tau_d}\left[g_I-a_I\right] \rho \\
%	J_{a_I} &:=& -\frac{a_I}{\tau_r} + \nu_I(t)\int_{a_I-\alpha_I}^{a_I} \rho(v_E,g_E,a_E,\eta_E,v_I,g_I,a_I',\eta_I,t)\,da_I' \\
%	J_{\eta_I} &:=& -\frac{1}{\tau_n} \eta_I\rho + \frac{1}{\tau_n}\frac{\partial^2 \rho}{\partial \eta_I^2} \\
The definitions of $g_{XY}$ in the LIF equations \eqref{e_lif}-\eqref{i_lif} result in a total conductance of $\gamma_{XY}g_Y$ on average.  Note that 
with a refractory period $\tau_{ref}>0$, the system of equations should also include a refractory probability density that we do not state here (see the work of Tranchina and colleagues 
\citet{ntslowi,HA,at06,ly_tranchina_09} for further details).

\subsection{Reduction theory to describe firing rate dynamics}\label{sec:red_theory}

We describe an insightful analytic reduction that captures how the range of excitatory firing rates change in different regimes.  We focus on only the excitatory neurons, which have fewer 
state variables if the inhibitory population is ignored or assumed to be known.  The problem with using the full p.d.f equations \eqref{full_fp_eqn}--\eqref{last_bc} is that the state variables are coupled, 
so we will formally assume that all of the state variables are known (given) except $v_E$, and solve for the steady-state firing rate as a function of the other state 
variables \citep{Ly_PrincDimRed_13,nlc_15}.  
Note that other authors have employed a similar approach using an adiabatic or slow variable approximation in the context of stochastic spiking neurons \citep{moreno3,nesse08}; very 
recently \citet{hertag14} used this approach formally (see their equation (25)).

{\bf We denote $r$ as the approximation to the excitatory firing rate(s) $\nu_E$}.  Assuming the other state variables are simply parameters, the deterministic firing rate of the equation 
$$\tau_m \frac{d v_E}{dt}  =  -v_E-q\tilde{g_{I}}(v_E-\mathcal{E}_I)-q\tilde{g_{E}}(v_E-\mathcal{E}_E)+\tilde{\eta_E}$$ 
is straightforward to calculate, and given by
\begin{equation}\label{r_det}
	r_0(q,\theta;\tilde{\vec{w}_E}) = \begin{cases} 
	0, & \text{if } \frac{q(\tilde{g_E}\mathcal{E}_E+\tilde{g_I}\mathcal{E}_I)+\tilde{\eta_E}}{1+q(\tilde{g_E}+\tilde{g_I})} \leq  \theta \\
	\frac{1+q(\tilde{g_E}+\tilde{g_I})}{\tau_m
	\log\left(\frac{q(\tilde{g_E}\mathcal{E}_E+\tilde{g_I}\mathcal{E}_I)+\tilde{\eta_E}}
		{q(\tilde{g_E}\mathcal{E}_E+\tilde{g_I}\mathcal{E}_I)+\tilde{\eta_E}-\theta(1+q(\tilde{g_E}+\tilde{g_I}))}\right)},  & \text{if }
		\frac{q(\tilde{g_E}\mathcal{E}_E+\tilde{g_I}\mathcal{E}_I)+\tilde{\eta_E}}{1+q(\tilde{g_E}+\tilde{g_I})} >  \theta 
	\end{cases}
\end{equation}
The argument of the left-hand side is written in this way because {\bf we assume the $q$ and $\theta$ values are the primary sources of heterogeneity}, rather than 
the external noise, finite size effects, random connectivity, etc.  
For exposition, we have absorbed the parameters and defined new variables with tildes: 
$\tilde{g_E}=\gamma_{ee} g_E$, $\tilde{g_I}=\gamma_{ei} g_I$, $\tilde{\eta_E}=\sigma_E \eta_E$.  
The approximation \eqref{r_det} ignores the refractory period, which is accounted for via a transformation rather than using the refractory probability density.  
The inverse of the firing rate is the time between spikes, so the refractory period can be added to the time between spikes to yield: $r_0/(1+\tau_{ref}r_0)$ as a 
simple approximation to the 
firing rate.  Finally, the given state variables are integrated against their marginal density to get:

\begin{equation}\label{thry_frate}
	r(q,\theta) = \mathbb{E}\left[\frac{r_0}{1+r_0\tau_{ref}}\right]=\int \frac{r_0}{1+r_0\tau_{ref}} \tilde{\rho}(\tilde{g_E},\tilde{g_I},\tilde{\eta_E}) \,d\tilde{\vec{w}_E} 
\end{equation}
There is a slight abuse of notation because the auxiliary variables $a_X$ effect the conductances but are not written in the previous equation; the emphasis is on how 
$(\tilde{g_E},\tilde{g_I},\tilde{\eta_E})$ directly effects $r$.  
Since the external noise is applied indiscriminately, $\tilde{\eta_E}$ is independent of the other variables and the marginal density factors into: 
$$\tilde{\rho}(\tilde{g_E},\tilde{g_I},\tilde{\eta_E})=\tilde{\rho}(\tilde{g_E},\tilde{g_I})\frac{e^{-(\tilde{\eta_E}/\sigma_E)^2}}{\sigma_E\sqrt{\pi}}.$$ 
However, $\tilde{\rho}(\tilde{g_E},\tilde{g_I})$ is still not analytically tractable, leading us to rely on Monte Carlo simulations to numerically estimate $\tilde{\rho}(\tilde{g_E},\tilde{g_I})$.

The reduction method \eqref{r_det}--\eqref{thry_frate} was implemented by relying on Monte Carlo simulations for $\tilde{\rho}(\tilde{g_E},\tilde{g_I})$, and using the same vectors 
$(\vec{q},\vec{\theta})$ in the LIF simulations.  Since $\tilde{\rho}(\tilde{g_E},\tilde{g_I})$ is the same for a given parameter set, the range of firing rates is theoretically captured 
by the different $(q_j,\theta_j)$ values in equation \eqref{thry_frate} (see blue curves in Figs. \ref{fig:2}--\ref{fig:4}).

\section{Results}

We consider a recurrently coupled stochastic spiking neural network.  Such networks have been ubiquitous in contemporary theoretical investigations.  
The class of networks considered here are widely used even though we do not include plasticity or detailed biophysical properties with different time scales.  We 
are interested in how the firing rate of the excitatory population changes as the level of heterogeneity is varied in different regimes.  Excitatory neurons are the focus because they are the 
predominate cells for propagating signals to different layers in the cortex.

\begin{figure}
	\includegraphics[width=\columnwidth]{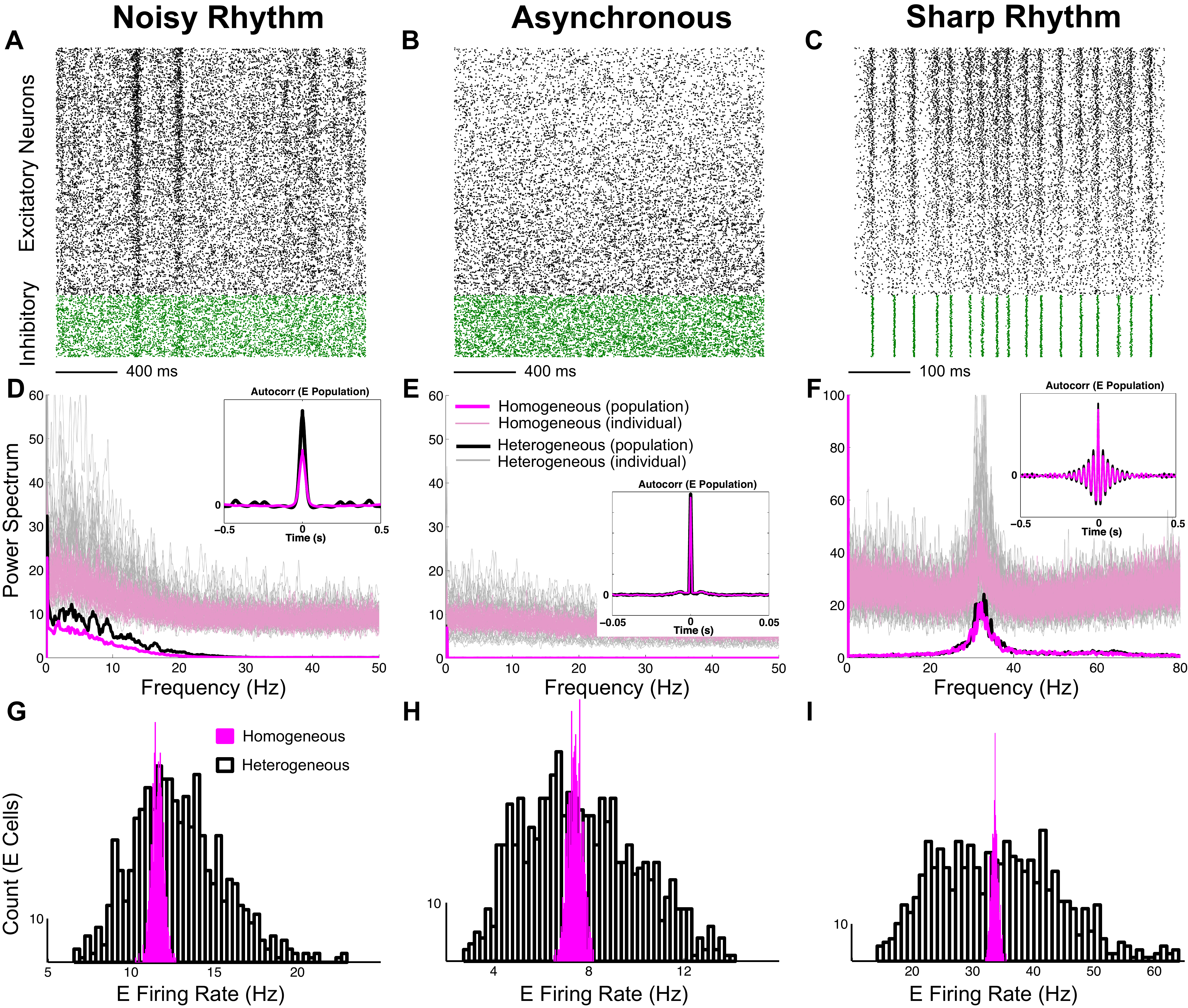}
\caption{The three regimes considered are: noisy rhythm (left column, A, D, G), asynchronous (middle column, B, E, H), sharp rhythm (right column, C, F, I).  
The top row (A--C) has representative raster plots of spikes for both excitatory (black dots) and inhibitory (green dots) neurons with both intrinsic and network heterogeneity, showing distinct behavior 
depending on the regime.  The middle row (D--F) shows the power spectrum of the excitatory population firing rate with both forms of heterogeneity (black) and homogeneous parameters (magenta); the 
power spectrum of the individual excitatory neurons are shown with thinner lines.  The inset of each panel shows the autocorrelation function of the excitatory population rate.  
The bottom row (G--I) has histograms of the average firing rate for each excitatory neuron, with the heterogeneous network naturally having a wider distribution.  The mean firing rate, minimum 
and maximum firing rates, and the range of the firing rates are displayed in Table \ref{tab:3}.  
The intrinsic and network heterogeneity parameters were selected independently with $\sigma_q=\sigma_\theta=1$ (see \eqref{het_sig1}--\eqref{het_sig2}).  
The simulations were performed for a single realization of 100\,s.}
\label{fig:1}       % Give a unique label
\end{figure}

Figure \ref{fig:1} highlights the different behaviors in three regimes considered.  Representative raster plots of spikes are shown in panels A--C for both excitatory and inhibitory cells, incorporating 
both forms of heterogeneity with $(\vec{q},\vec{\theta})$ chosen independently and with $\sigma_q=\sigma_\theta=1$.  
In panels D--F, the power spectrum of the excitatory population firing rate for both 
heterogeneous (black) and homogeneous parameters (magenta) are shown; the thinner lines are the power spectrums of the individual excitatory neurons.  The inset shows the autocorrelation 
function of the (E) population firing rate.  The autocorrelation function of a stochastic process $R(t)$ (e.g., excitatory population firing rate) is:
\begin{equation}\label{aut_defn}
	A(t) = \mathbb{E}_\tau \left[ R(\tau)R(t-\tau) \right] - \mathbb{E}_\tau[R(\tau)]^2
\end{equation}
and the power spectrum is:
\begin{equation}\label{powsp_defn}
	P(\omega) = \Big\vert \int A(t) e^{-i 2\pi \omega t}\,dt \Big\vert.
\end{equation}
These quantities illustrate the different dynamics of each neural network.  
Notice that the power spectrum of the individual neurons is consistently larger in value than the power spectrum of the population firing rate.  
This is because the population firing rate is averaged (smoothed) so that 
$P(\omega)\to0$ for large frequencies, whereas the spike train of the individual neurons is not smooth, consisting of 0's or $\delta's\approx 1/(\Delta t)$ that will always 
yield $P_j>0$ as long as $\omega<1/(\Delta t)$ (the numerical limit because of the discretization).  Specifically, as $\omega\to 1/(\Delta t)$, $P_j$ measures the average power of the spike 
train in a single time bin and thus converges to the firing rate of the individual neuron.  
The bottom row (panels G--I) show the distribution of the excitatory firing rates for each individual neuron, averaged over time (simulations in Figure \ref{fig:1} were performed for 100\,s).  
The heterogeneous population naturally has a wider distribution compared to the homogeneous regime.  At the population level, 
there are only minor differences between the homogeneous and heterogeneous regimes; indeed, the average firing rates (not the overall distribution), power spectrums, and 
autocorrelation functions are very similar.  
Thus enabling a systematic assessment of how intrinsic and network heterogeneity effect the spiking network, avoiding the complication of regime changes due to heterogeneity.  
Although there has been interesting work showing how heterogeneity can induce rhythms from asynchrony (i.e., bifurcations) \citep{hermann12,mejias12}, we do not directly address such 
dynamics here.  Our study focuses on comparing the firing rate heterogeneity modulation {\it within} specific regimes.

The neurons in all regimes considered are all excitable and receive external colored noise, resulting in irregular spiking.  A common measure of variability is the Fano factor, defined as the variance of spike counts 
divided by the mean of spike counts in a time window.  For all of the networks considered in this paper (e.g., Fig. \ref{fig:1}), the Fano factor of the E population spike counts is often greater than 
1 and is at least 0.8 for heterogeneous and homogeneous networks, across all regimes, and for time windows ranging from 2 to 50\,ms (not shown).

\begin{table}
\caption{Firing Rate Values in Figure \ref{fig:1} G--I.  The inhibitory firing rates are not shown in Figure \ref{fig:1}.}
\label{tab:3}       % Give a unique label
\begin{tabular}{lllllll}
\hline\noalign{\smallskip}
{\bf Regime:} & Mean $\nu$ (Hz) & $[\nu_{min},\nu_{max}]$ (Hz) & Range of $\nu$ (Hz) \\
\noalign{\smallskip}\hline\noalign{\smallskip}
{\bf Noisy Rhythm } (Heterog.): E cells & 12.8 & [6.6, 23] &  16.4 \\
{\bf Noisy Rhythm } (Homog.): E cells & 11.6 & [10.3, 12.7] &  2.5 \\
{\bf Asynchronous } (Heterog.): E cells & 7.7 & [2.7, 14.2] &  11.4 \\
{\bf Asynchronous } (Homog.): E cells & 7.4 & [6.5, 8.2] &  1.7 \\
{\bf Sharp Rhythm } (Heterog.): E cells & 34.1 & [14.2, 64] &  49.8 \\
{\bf Sharp Rhythm } (Homog.): E cells & 33.6 & [32.1, 35.5] &  3.5 \\
{\bf Noisy Rhythm } (Heterog.): I cells & 17 & [15.9, 18.7] &  2.8 \\
{\bf Noisy Rhythm } (Homog.): I cells & 16.1 & [14.8, 17.4] &  2.6 \\
{\bf Asynchronous } (Heterog.): I cells & 19.7 & [18.5, 21.2] &  2.7 \\
{\bf Asynchronous }: Homog. I cells & 19.6 & [17.9, 20.9] &  3 \\
{\bf Sharp Rhythm } (Heterog.): I cells & 26.8 & [17.4, 32.4] &  15 \\
{\bf Sharp Rhythm } (Homog.): I cells & 26 & [24.3, 27.9] &  3.6 \\
\noalign{\smallskip}\hline
\end{tabular}
\end{table}

\subsection{PDF framework captures firing rate modulation with heterogeneity}\label{sec:pdf_het}

\begin{figure}
	\includegraphics[width=0.9\columnwidth]{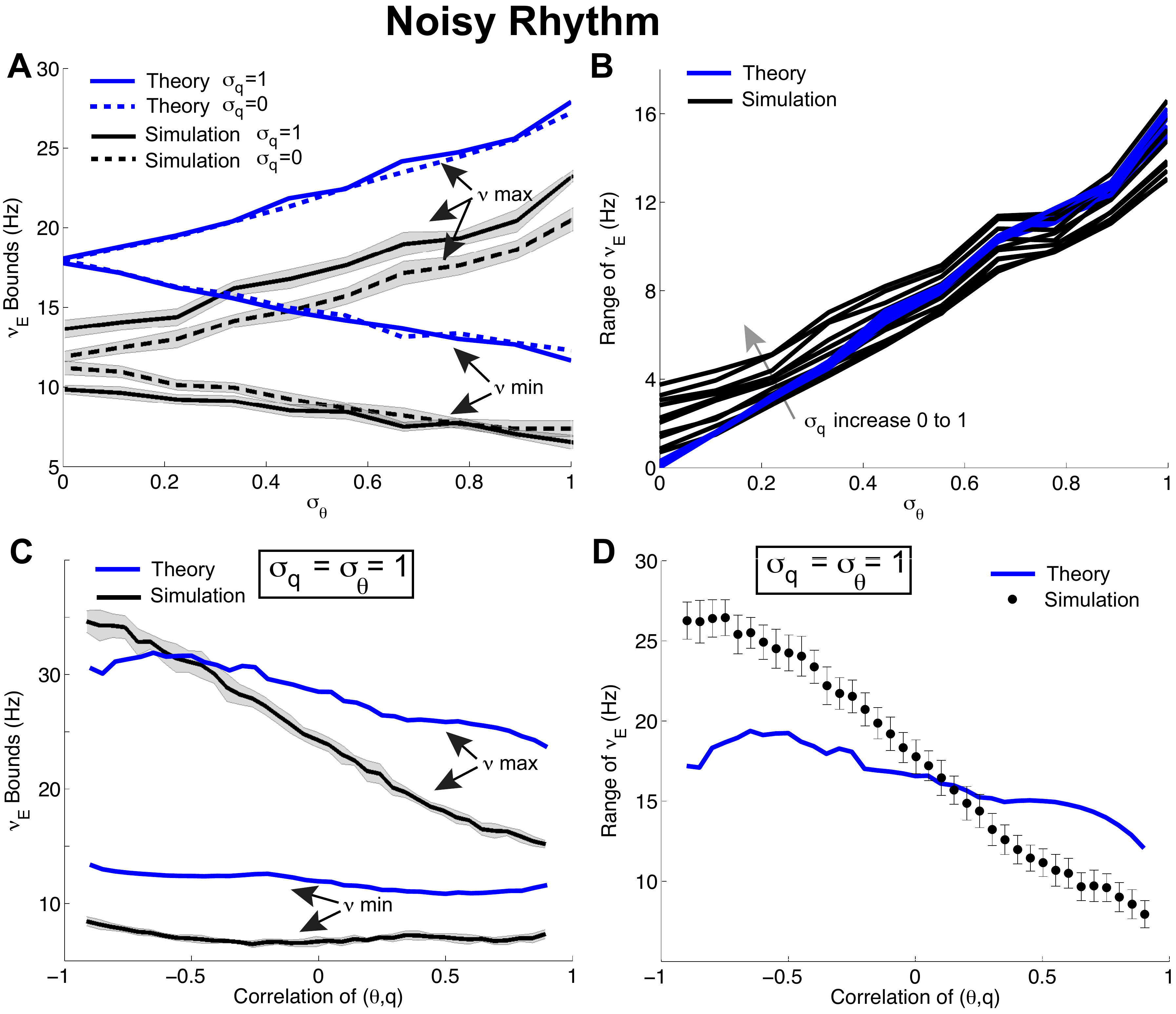}
\caption{Noisy Rhythm regime: excitatory firing rates modulation with changes in intrinsic and network heterogeneity.  A)--B) changing the level of intrinsic $\sigma_\theta$ and 
network $\sigma_q$ heterogeneity by independently drawing $(\vec{q},\vec{\theta})$ (see \eqref{het_sig1}--\eqref{het_sig2}).  A) The minimum and maximum firing rates are shown for 
the homogeneous network ($\sigma_q=\sigma_\theta=0$, dash curves) and completely heterogeneous ($\sigma_q=\sigma_\theta=1$, solid curves) network.  The simulation curves 
(black curves, dash and solid) are Monte Carlo simulations of equations \eqref{e_lif}--\eqref{i_lif} (shaded regions denote standard deviations).  Here, the {\bf theory} uses 
a combination of dimensionally reduced PDF functions and Monte Carlo simulations (see \eqref{r_det}--\eqref{thry_frate}).  Although the theory does not quantitatively match the 
extreme values of the firing rates (A), it captures the trend of the firing rate range (maximum minus minimum) in panel B); there, each curve is for a fixed value of 
$\sigma_q$ ranging from 0 to 1, as $\sigma_\theta$ ranges between 0 and 1.  Unsurprisingly, as heterogeneity increases so does the firing rate range.  
The firing rates and the range vary appreciably over an order of magnitude.  C)--D) changing the correlation $\varrho$ between $(\vec{q},\vec{\theta})$ with $\sigma_q=\sigma_\theta=1$.  
C) the theory (blue curve) does not quantitatively capture the actual firing rates as $\varrho$ varies between (-1,1), but they are comparable.  D) the theory (solid) 
captures the trend in the simulated range of the firing rates (dots) as $\varrho$ varies between (-1,1).  As $\varrho$ increases, the range of firing rates tends to 
decrease.  Gray shaded regions in A and C are an estimate of the standard deviation about the sample mean of the Monte Carlo simulations (100\,s simulation for each realization, 10 realizations total); error bars in 
D are estimates of the standard deviation about the sample mean of the range (see last paragraph of Section \ref{sec:lif} for details).  Shaded regions are omitted in B for readability.}
\label{fig:2}       % Give a unique label
\end{figure}

The level of intrinsic and network heterogeneity were modulated in the recurrently coupled networks in the two ways previously described: i) independently choose the vectors $(\vec{q},\vec{\theta})$ 
with a prescribed level of heterogeneity determined by $\sigma_q$ and $\sigma_\theta$ (see \eqref{het_sig1}--\eqref{het_sig2}), and ii) for fixed values of $\sigma_{q/\theta}$, change the correlation between 
$(\vec{q},\vec{\theta})$.  %We are interested in how the firing rates of the entire excitatory neuron population changes with varying levels of heterogeneity.  
In the noisy rhythm regime, Figure \ref{fig:2}A shows the minimum and maximum excitatory firing rates for fixed values of $\sigma_q=0$ and 1, while $\sigma_\theta$ varies between 0 and 1 (black 
curves; dashed and solid).  Not surprisingly, as $\sigma_\theta$ (or $\sigma_q$) increased, so does the range of (excitatory) firing rates with the minimum decreasing and maximum increasing.  
Note that we chose to plot the curves for fixed $\sigma_q$; the results also hold with fixed $\sigma_\theta$ and $\sigma_q$ on the x-axis (not shown).  
Figure \ref{fig:2}B shows the range of the firing rates (maximum minus the minimum) rather than the raw firing rate values; it is apparent that more intrinsic and/or network heterogeneity leads 
to more firing rate heterogeneity.  The reduction theory 
described in section \ref{sec:red_theory} (based on both probability density equations and Monte Carlo simulations), and in particular equations \eqref{r_det}--\eqref{thry_frate} for the 
approximation of the firing rates, are shown in the blue colored curves.  In Figure \ref{fig:2}A, the 
theory does not provide a good quantitative match.  This could be due to a variety of reasons: the 4 assumptions in section \ref{sec:hd_pdf} are violated, the reduction method is known to 
be inaccurate compared 
to both the full PDF and Monte Carlo simulations.  Fortunately, the reduction theory is able to capture the increase in the firing rate range in Figure \ref{fig:2}B, where $\sigma_{q,\theta}\in[0,1]$.  
This result is indeed fortuitous given the inaccuracies of the PDF theory in capturing the raw firing rates.

Figure \ref{fig:2}C shows the minimum and maximum firing rate with ample heterogeneity ($\sigma_q=\sigma_\theta=1$, the most we considered), but the correlation between $\vec{q}$ and $\vec{\theta}$ 
($\varrho(\vec{q},\vec{\theta})$) varied between (-1,1).  The comparison of the reduction theory \eqref{r_det}--\eqref{thry_frate} (blue curves) and the simulations (black curves) is not accurate 
(as in Fig. \ref{fig:2}A), but does qualitatively capture the trend in the range of the firing rates (Fig. \ref{fig:2}D).  The range of firing rates tends to decrease as $\varrho(\vec{q},\vec{\theta})$ increases.  Note 
that the range of firing rates can be very large and very small depending on $\varrho$, even though $\sigma_q=\sigma_\theta=1$; in fact, the ranges of firing rates are comparable to varying 
the overall level of heterogeneity: $\sigma_{q/\theta}\in[0,1]$ (Fig. \ref{fig:2}A--B).  In other words, a particular range of excitatory firing rates can arise from different levels of intrinsic and network 
heterogeneity, {\it depending on their relationship}.

\begin{figure}
	\includegraphics[width=0.9\columnwidth]{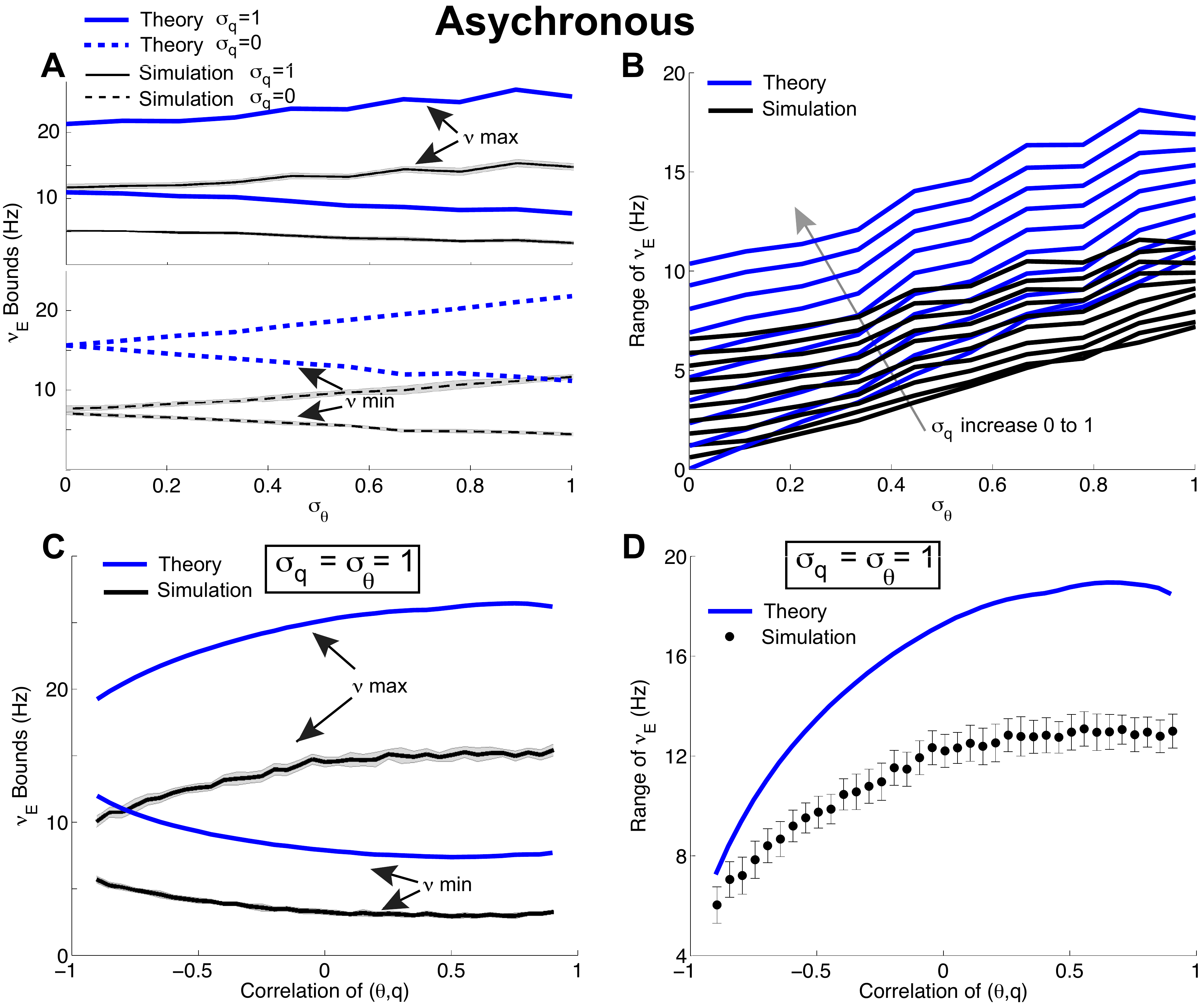}
\caption{Asynchronous regime: excitatory firing rates modulation with changes in intrinsic and network heterogeneity [similar to Figure \ref{fig:2}] A)--B) changing the level of 
intrinsic $\sigma_\theta$ and 
network $\sigma_q$ heterogeneity by independently drawing $(\vec{q},\vec{\theta})$ (see \eqref{het_sig1}--\eqref{het_sig2}).  A) The minimum and maximum firing rates are shown for 
the completely heterogeneous ($\sigma_q=\sigma_\theta=1$, solid curves) network in the top panel and 
the homogeneous network ($\sigma_q=\sigma_\theta=0$, dash curves) in the bottom panel.  The simulation curves 
(black curves, dash and solid) are Monte Carlo simulations of equations \eqref{e_lif}--\eqref{i_lif}.  Here, the {\bf theory} uses 
a combination of dimensionally reduced PDF functions and Monte Carlo simulations (see \eqref{r_det}--\eqref{thry_frate}).  Although the theory does not quantitatively match the 
extreme values of the firing rates (A), it captures the trend of the firing rate range (maximum minus minimum) in panel B); there, each curve is for a fixed value of 
$\sigma_q$ ranging from 0 to 1, as $\sigma_\theta$ ranges between 0 and 1.  Unsurprisingly, as heterogeneity increases so does the firing rate range.  
C)--D) changing the correlation $\varrho$ between $(\vec{q},\vec{\theta})$ with $\sigma_q=\sigma_\theta=1$.  
C) the theory (blue curve) does not quantitatively capture the actual firing rates as $\varrho$ varies between (-1,1).  D) the theory (solid) 
captures the trend in the simulated range of the firing rates (dots) as $\varrho$ varies between (-1,1).  As $\varrho$ increases, the range of firing rates tends to 
increase.  Gray shaded regions in A and C are an estimate of the standard deviation about the sample mean of the Monte Carlo simulations (100\,s simulation for each realization, 10 realizations total); error bars in 
D are estimates of the standard deviation about the sample mean of the range (see last paragraph of Section \ref{sec:lif} for details).  Shaded regions are omitted in B for readability.}
\label{fig:3}       % Give a unique label
\end{figure}

\begin{figure}
	\includegraphics[width=0.9\columnwidth]{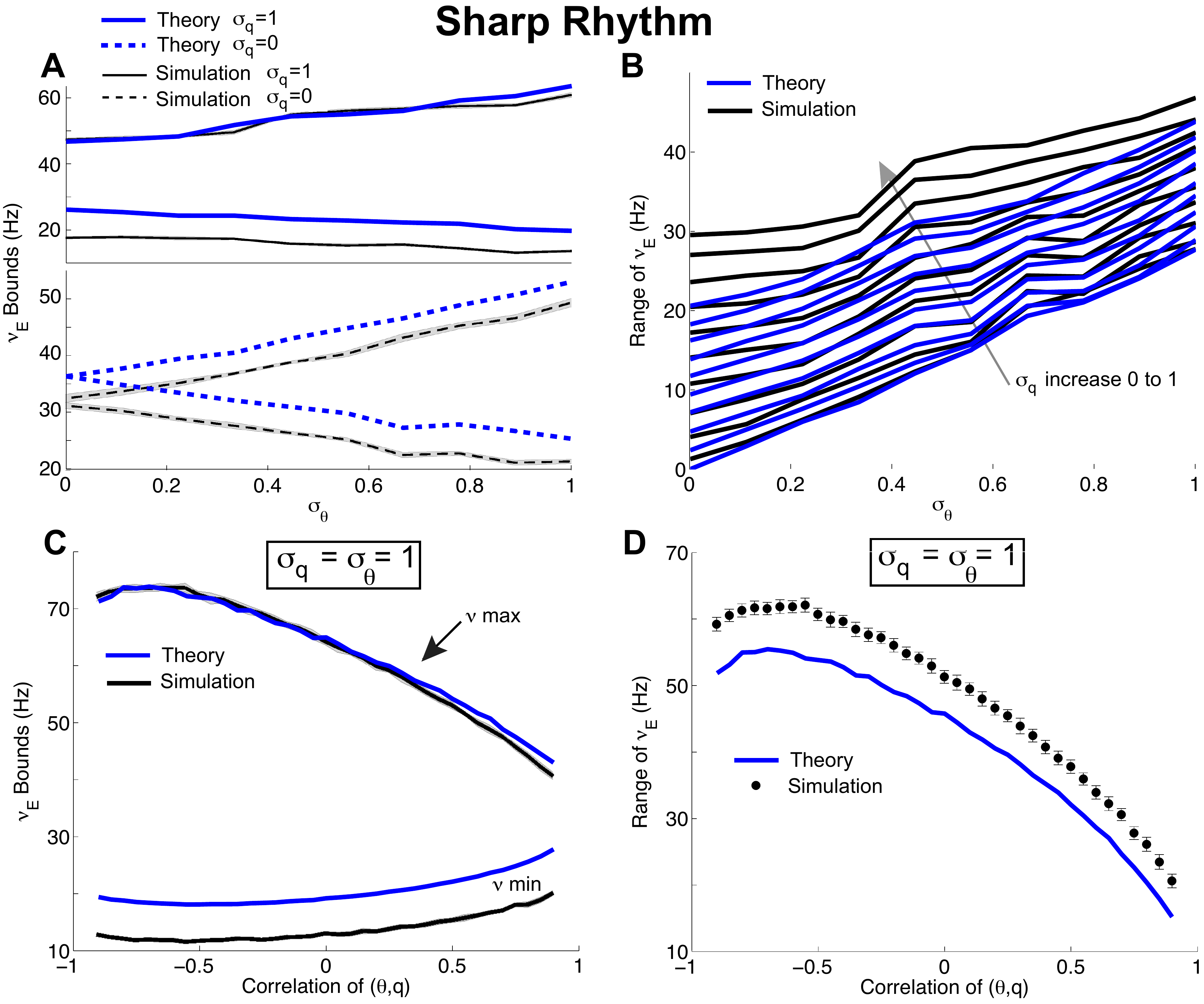}
\caption{Sharp Rhythm regime: excitatory firing rates modulation with changes in intrinsic and network heterogeneity [similar to Figures \ref{fig:2}--\ref{fig:3}] 
A)--B) changing the level of intrinsic $\sigma_\theta$ and 
network $\sigma_q$ heterogeneity by independently drawing $(\vec{q},\vec{\theta})$ (see \eqref{het_sig1}--\eqref{het_sig2}).  A) The minimum and maximum firing rates are shown for 
the completely heterogeneous ($\sigma_q=\sigma_\theta=1$, solid curves) network in the top panel and 
the homogeneous network ($\sigma_q=\sigma_\theta=0$, dash curves) in the bottom panel.  The simulation curves 
(black curves, dash and solid) are Monte Carlo simulations of equations \eqref{e_lif}--\eqref{i_lif}.  Here, the {\bf theory} uses 
a combination of dimensionally reduced PDF functions and Monte Carlo simulations (see \eqref{r_det}--\eqref{thry_frate}).  Although the theory does not quantitatively match the 
extreme values of the firing rates (A), it captures the trend of the firing rate range (maximum minus minimum) in panel B); there, each curve is for a fixed value of 
$\sigma_q$ ranging from 0 to 1, as $\sigma_\theta$ ranges between 0 and 1.  Unsurprisingly, as heterogeneity increases so does the firing rate range.  
C)--D) changing the correlation $\varrho$ between $(\vec{q},\vec{\theta})$ with $\sigma_q=\sigma_\theta=1$.  
C) again, the theory (blue curve) does not quantitatively capture the actual firing rates as $\varrho$ varies between (-1,1).  D) the theory (solid) 
captures the trend in the simulated range of the firing rates (dots) as $\varrho$ varies between (-1,1).  As $\varrho$ increases, the range of firing rates tends to 
decrease.  Gray shaded regions in A and C are an estimate of the standard deviation about the sample mean of the Monte Carlo simulations (100\,s simulation for each realization, 10 realizations total); error bars in 
D are estimates of the standard deviation about the sample mean of the range (see last paragraph of Section \ref{sec:lif} for details).  Shaded regions are omitted in B for readability.}
\label{fig:4}       % Give a unique label
\end{figure}

Similar comparisons are made for the two other regimes in Figures \ref{fig:3} and \ref{fig:4}.  In the asynchronous regime, as the the heterogeneity parameters are selected 
independently (Fig. \ref{fig:3}A--B), we again see that more heterogeneity leads to a wider range of firing rates.  Figure \ref{fig:3}A shows $\sigma_q=0$ and 1 split into two panels so it is easier to compare 
the theory \eqref{r_det}--\eqref{thry_frate} and simulations \eqref{e_lif}--\eqref{i_lif}.  The quantitative match is not good (Fig. \ref{fig:3}A) as expected given the previous figure, 
but the trend is captured (Fig. \ref{fig:3}B, where $\sigma_{q,\theta}\in[0,1]$).  For the bottom row, we fix 
$\sigma_q=\sigma_\theta=1$ and let the correlation $\varrho$ between intrinsic and network heterogeneity vary between (-1,1).  Notice that the range of firing rates changes in a different way 
(Fig. \ref{fig:3}D).  Here, as $\varrho$ increases, the range of firing rates {\it increases} in contrast to before where it decreased (Fig. \ref{fig:2}D).  
Moreover, we see the range of firing rates change by a factor of $\sim$3, which interestingly is comparable to 
the firing rate range values when varying $\sigma_{q/\theta}$ independently (Fig. \ref{fig:3}A--B).

In Figure \ref{fig:4}, the sharp rhythm regime shows similar characteristics to Figure \ref{fig:2}, except for the following.  The excitatory firing rate range is more sensitive to varying the 
heterogeneity parameters, and we see that the range of firing rates takes on much larger values.  This is apparent both when $\sigma_q$ and $\sigma_\theta$ vary independently 
(Fig. \ref{fig:4}A--B) and when the correlation between $\vec{q}$ and $\vec{\theta}$ changes with $\sigma_q=\sigma_\theta=1$ (Fig. \ref{fig:4}C--D).  
The firing rate range changes by almost an order of magnitude.  The 
other interesting thing about this regime is that the reduction theory (blue) matches the simulations much better than the other two regimes (Fig. \ref{fig:2}--\ref{fig:3}).  The match is particularly 
good for the maximum firing rate.  Similar to the noisy rhythm regime, we see that as $\varrho$ increases, the range of firing rates decreases dramatically (Fig. \ref{fig:4}D).  
One striking observation is that the PDF theory captures the firing rates much better in the sharp rhythm regime (Fig. \ref{fig:4}C--D) than the other two regimes (Fig. \ref{fig:2}C--D, 
Fig. \ref{fig:3}C--D).  Although the underlying reason for this is difficult to determine exactly, a plausible explanation is that the overall higher firing rates and less noise in the sharp rhythm regime 
are consistent with the assumptions of the approximation in equation \eqref{r_det}.

The observation that more intrinsic and (uncorrelated) network heterogeneity leads to a wider range of firing rates (Figs. \ref{fig:2}--\ref{fig:4}, panels A and B) is expected and does not require further 
analytical insight.  However, changing the correlation between $\vec{q}$ and $\vec{\theta}$ for fixed values of $\sigma_q$ and $\sigma_\theta$ results in enlarged or diminished ranges of firing rates, 
depending on the regime (Figs. \ref{fig:2}--\ref{fig:4}, panels C and D).  {\bf How can these observations be reconciled?}  
The next two sections provide further analysis to deeply understand these phenomena.

\subsection{Analytic description of heterogeneous firing rate range in rhythmic networks}\label{sec:thry_h1}

For many regimes and the different types of heterogeneity, the reduction method in section \ref{sec:red_theory} does qualitatively capture the modulation of the range of firing rates, as shown in 
Figures \ref{fig:2}--\ref{fig:4}, panels B and D.  This observation is the 
motivation for further analysis of equations \eqref{r_det}--\eqref{thry_frate} that ultimately yields simple analytic formulas to account for how the firing rate ranges are effected by the relationship between intrinsic and 
networks heterogeneity.  
In sections \ref{sec:thry_h1} and \ref{sec:thry_h2}, we use the following variable substitutions to facilitate exposition of the analysis:
\begin{eqnarray}
	x_0 & := &  	\tilde{g_E}+\tilde{g_I} \label{x_def1} \\
	x_1 & := & 	\tilde{g_E}\mathcal{E}_E+\tilde{g_I}\mathcal{E}_I \label{x_def2}
\end{eqnarray}

When the coupling parameters yield rhythmic firing (i.e., power spectrum of the population firing rate is not flat), the net synaptic input is large on average (averaged over time 
and across excitatory neurons) and is much larger than when the 
network is in an asynchronous regime.  Thus, we consider the large firing rate limit in the reduced theoretical description \eqref{r_det}--\eqref{thry_frate}.  Furthermore, we ignore the effects of the 
refractory period $\tau_{ref}$ \footnote{Although ignoring the refractory period could be problematic for large firing rates, we emphasize that the value of our analysis is not in quantitative matching of simulations but rather for a deeper understanding 
of how network attributes effect the outputs.  A similar calculation has been performed with the refractory period (not shown), but the asymptotic formulas are not insightful.}, 
external noise $\tilde{\eta_E}$, and focus on the formula for the deterministic firing rate \eqref{r_det}:
\begin{equation}\label{rdet_osc}
	\tau_m r_0(q,\theta) = \frac{1+qx_0}{\log\left(\frac{qx_1}
		{qx_1-\theta(1+qx_0)} \right)}
\end{equation}
We assume the random state variables are parameters just like in the reduced description, which will enable us to focus on $(q,\theta)$ and determine how these two parameters effect the firing rate range.  
The two vectors $(\vec{q},\vec{\theta})$ are the main source of the firing rate heterogeneity.  
The large firing rate regime is captured by a series expansion of $\log()$ around 1.  Standard asymptotic calculations enable equation \eqref{rdet_osc} to be re-written as:
\begin{eqnarray}
	\tau_m r_0(q,\theta) &=& \frac{1+qx_0}{\log\left(\frac{qx_1}
		{qx_1-\theta(1+qx_0)} \right)} \\
		&=& \frac{q}{\theta}x_1-\frac{1}{2}(1+qx_0) \nonumber \\
		& &-\frac{(1+qx_0)^2\theta}{12[qx_1-\theta(1+qx_0)]}+O\left(z^2(1+qx_0)\right) \label{rhyth_expan} \\
	 \hbox{where } z &:=& \theta \frac{1+qx_0}{qx_1-\theta(1+qx_0)} \label{z_defn}
\end{eqnarray}
The modulation of the firing rate heterogeneity can be understood simply by the dominant term (i.e., first term):
\begin{equation}\label{rhyth_form}
	\frac{q}{\theta}x_1;
\end{equation}
specifically, the fraction $q/\theta$ is key because the second factor $x_1$ does not vary much as the correlation 
between $\vec{q}$ and $\vec{\theta}$ changes.  

In the rhythmic firing regime, the modulation of the range of firing rates can be understood simply with range of values given by the fraction $q_j/\theta_j$.  
Specifically, the ratio $q_j\frac{1}{\theta_j}$ yields $N_e$ values, and the range of these values for a particular parameter is indicative of the relative range of firing rates.  
In this regime (rhythmic firing) when $\varrho(\vec{q},\vec{\theta})<0$, 
the extreme values consists of: i) larger $q_j$ values that tend to occur with smaller $\theta_j$ values, resulting in an amplification (and relatively larger) $q\frac{1}{\theta}$ for the upper 
range of values, and ii) smaller $q_j$ values and larger $\theta_j$, resulting in an overall smaller (small times small) values to account for the lower range of $q\frac{1}{\theta}$ values.  
When $\varrho(\vec{q},\vec{\theta})>0$, similar reasoning applies to larger $q_j$ values tending to occur with larger $\theta_j$ values (large times small) and smaller values (small times large), 
resulting in a diminished range of $q\frac{1}{\theta}$ values than when $\varrho>0$.  Note that the mean of $q_j/\theta_j$ (across the $N_e$ population) is approximately constant as 
$\varrho$ varies (not shown).

\begin{figure}
	\includegraphics[width=.95\columnwidth]{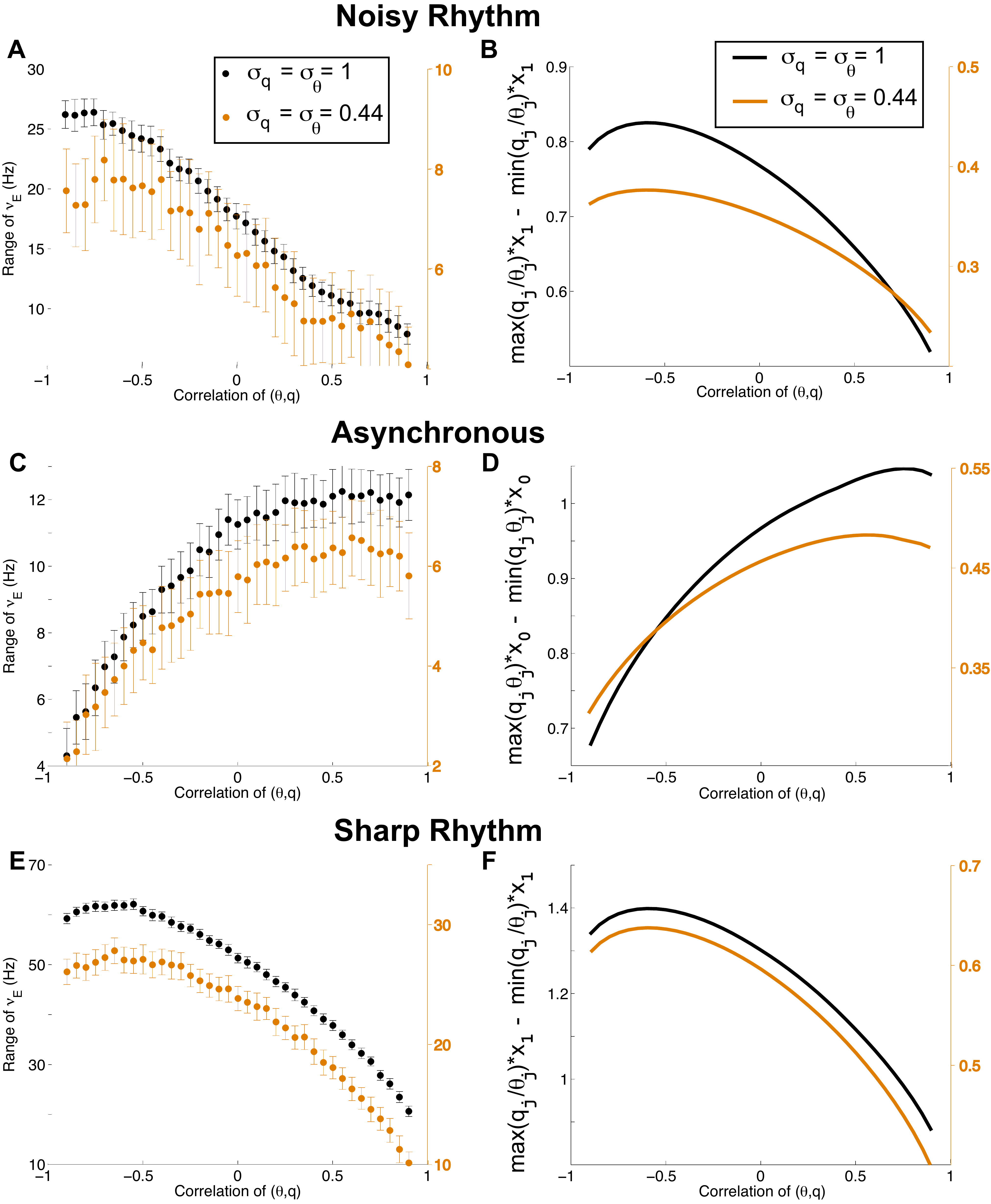}
\caption{Analytic description of excitatory firing rate ranges.  Here, the firing rate ranges are for a fixed level of heterogeneity $\sigma_q=\sigma_\theta$ and 
different correlation $\varrho(\vec{q},\vec{\theta})$ values.  A)--B) Noisy Rhythm regime: 
the range of firing rates for the completely heterogeneous ($\sigma_q=\sigma_\theta=1$, black dots) network and with less heterogeneity 
($\sigma_q=\sigma_\theta=0.44$, dark orange dots) are plotted as the correlation varies on the same axis (see left and right vertical axes for respective scales).  
B) The theory in section \ref{sec:thry_h1} provides an analytic description (see \eqref{rhyth_form}) for how the correlation of intrinsic and network heterogeneity lead to 
relatively different firing rate ranges in this regime.  As the correlation $\varrho$ increases, the range of firing rates tends to decrease.  
C)--D) Asynchronous regime: similar to A)--B), but the reduction theory in D) is in section \ref{sec:thry_h2}, again providing an analytic description (see \eqref{asynch_form}) for 
how the range of firing rates increases as the correlation $\varrho$ increases.  E)--F) Sharp Rhythm regime: similar to A)--B), using the same analytic description.  The 
analytic descriptions in the right column (B, D, F) use $x_0$ and $x_1$ (see \eqref{x_def1}--\eqref{x_def2}) values obtained from $\varrho=0$.  
Error bars in A, C, and E are estimates of the standard deviation about the sample mean of the range (see last paragraph of Section \ref{sec:lif} for details).}
\label{fig:5}       % Give a unique label
\end{figure}

A side-by-side comparison shows how similar the dynamics 
are (Fig. \ref{fig:5}A--B, E--F) and is validation that the range of excitatory firing rates is driven by this factor $(\vec{q}/\vec{\theta}$).  
Figure \ref{fig:5} compares the excitatory firing rate range of the LIF simulations (left column) to the analytic descriptions 
(right column, sections \ref{sec:thry_h1}--\ref{sec:thry_h2}) as a function of $\varrho(\vec{q},\vec{\theta})$ for two fixed levels of heterogeneity: $\sigma_q=\sigma_\theta=1$ (black) 
and $\sigma_q=\sigma_\theta=0.44$ (dark orange).  
Of course we do not expect precise quantitative matching between the analytic description (Fig. \ref{fig:5} right column) and the actual network simulation (Fig. \ref{fig:5} left column) 
because the analysis had many assumptions meant to highlight a proof of principle.  Nevertheless, the theory is quite descriptive and captures the general trend (the values of 
equation \eqref{rhyth_form} are shown in Fig. \ref{fig:5}B,F for completeness).  
Therefore, in the rhythmic firing regime, $q/\theta$ compactly describes how when intrinsic and network heterogeneity are anti-correlated the range of firing rates is larger than when they are correlated.  

\subsection{Analytic description of heterogeneous firing rate range in asynchronous networks}\label{sec:thry_h2}

For the asynchronous regime, the range of firing rates actually increases as the correlation between $\vec{\theta}$ and $\vec{q}$ increases (Fig. \ref{fig:3}D), 
which is the opposite trend compared to when the network is firing rhythmically (Fig. \ref{fig:2}D and \ref{fig:4}D).  This section provides an analytic description for this phenomena.

In contrast to when the population firing rate is rhythmic, the asynchronous regime (i.e., power spectrum of population firing rate is flat) has much less net synaptic input on average.  
Therefore, we cannot ignore the noise variable $\tilde{\eta_E}$ and must consider a different regime than in the previous section.  Here, the refractory period $\tau_{ref}$ is ignored, and all of the random 
variables are again assumed to be parameters {\it except} the noise variable $\tilde{\eta_E}$ because of how crucial it is for firing in this regime.  The formula 
for the reduced firing rate is (cf. equations \eqref{thry_frate} and \eqref{rdet_osc}):

\begin{equation}\label{r_asyn}
	\tau_m r_0(q,\theta) = \frac{1+qx_0}{\sigma_E\sqrt{\pi}} \int 
	\left\lceil \frac{1}{\log\left(\frac{qx_1+\tilde{\eta_E}}
		{qx_1+\tilde{\eta_E}-\theta(1+qx_0)} \right)} \right\rceil^+ e^{-\tilde{\eta_E}^2/\sigma_E}\,d\tilde{\eta_E}
\end{equation}
where $\lceil \,\, \rceil^+$ represents thresholding: if $\tilde{\eta_E}\leq \theta(1+qx_0)-qx_1$ it is 0, otherwise it is $1/\log(\cdot)$.  
This equation can be re-written by substituting $\lceil 1/\log(\cdot) \rceil^+$ with an integral, and interchanging the order of integration:
\begin{eqnarray}
	\tau_m r_0(q,\theta) &=& \frac{1+qx_0}{\sigma_E\sqrt{\pi}} \int_{-\infty}^\infty 
	\int_{qx_1+\tilde{\eta_E}-\theta(1+qx_0)}^{qx_1+\tilde{\eta_E}} 
	M(y)\,dy e^{-\tilde{\eta_E}^2/\sigma_E}\,d\tilde{\eta_E} \nonumber \\
&=& \frac{1+qx_0}{\sigma_E\sqrt{\pi}} \int_{-\infty}^{\infty} \int_{y-qx_1}^{y-qx_1+\theta(1+qx_0)} e^{-\tilde{\eta_E}^2/\sigma_E}\,d\tilde{\eta_E}  M(y)\,dy
\end{eqnarray}
where $M(y)$ is an antiderivative of $\lceil 1/\log(\cdot) \rceil^+$.  
In the asynchronous firing regime, the modulation of the range of firing rates can be understood simply with the last term of upper limit of the inner integral: 
\begin{equation}\label{asynch_form}
	\theta(1+qx_0)
\end{equation}
because the other pieces of $r_0$ do not vary much across the $N_e$ neurons.  Furthermore, since the focus is on understanding the range of firing rates as determined by the 
correlation of $(\vec{q},\vec{\theta})$, then the product: $\theta q$ 
%Recall that the heterogeneity is predominately controlled by $(\vec{q},\vec{\theta})$.  
is what determines the relative change in the range (like before, $x_0$ does not change much across the excitatory neurons).  
This is in sharp contrast to equation \eqref{rhyth_form}, where the intrinsic heterogeneity $\theta$ divides the network heterogeneity $q$.  
Figure \ref{fig:5}C--D shows that the analytic description \eqref{asynch_form} clearly captures the trend in the range of firing rates as $\varrho$ changes.

These formulas \eqref{rhyth_form}, \eqref{asynch_form}, clearly show how the two forms of heterogeneity $(\vec{q},\vec{\theta})$ effect one another to yield the different trends in the range of 
firing rates in different regimes.  Simply put, in the rhythmic regime, the relationship $q/\theta$ determines how the range of the firing rates change while in the asynchronous regime, $q\theta$ determines 
this; note that plotting $q/\theta$ and $q\theta$ without $x_0$ and $x_1$ does not appreciably change the shape of the curves in Figure \ref{fig:5} (not shown).

The key to our analysis was to focus on the main sources of heterogeneity, and therefore the firing rate heterogeneity, in a proper framework (probability density functions) to describe the essence of how the 
firing rate range changes.  This analysis was successful partly because the heterogeneity $(q,\theta)$ drove the changes in the firing rate range, as opposed to other factors such as external noise, etc., and our 
analysis centered on these parameters.

\section{Discussion}\label{discuss}

We studied how two forms of heterogeneity: intrinsic and network, effect the (excitatory) firing rate distribution of a recurrently coupled stochastic network of leaky integrate-and-fire neurons.  
Since the relationship between 
intrinsic and network heterogeneity is not known (to the best of our knowledge), we systematically varied the relationship or correlation to assess the effects on the network in different regimes.  
This work showed how the firing rate range changes with the correlation of intrinsic and network heterogeneity: in the rhythmic or oscillatory regime, the firing rate range tends to decrease with 
increasing correlation (i.e., when 
larger firing thresholds tend to have larger synaptic input amplification), while the opposite trend is observed in the asynchronous regime.  These observations were captured by the analytic descriptions in 
equations \eqref{rhyth_form} and \eqref{asynch_form}.  We also found that the firing rate ranges can be relatively large or small depending on the correlation between intrinsic and network 
heterogeneity, thus the overall level of heterogeneity can be mitigated or amplified depending on this relationship.  If the relationship between intrinsic and network heterogeneity could be measured in a 
cortical network, these results would enable predictions for the range of response heterogeneity.  Although we chose to analyze two specific forms of heterogeneity in a theoretical model, connections to 
experimental recordings of heterogeneity and firing rates may be possible even with related neural attributes (i.e., the membrane time constant as a proxy for firing threshold, or relating network 
connectivity to input variability).  
Also, the framework presented here could in principle be adapted to other heterogeneous neural attributes that would naturally require augmentations.

\citet{marder11} showed how combining intrinsic and synaptic conductance heterogeneity could lead to similar outputs, depending on their relationship.  Their work was naturally different than the 
work here: different parameters were varied and the underlying neuron model had ionic currents and specific circuitry motivated by the crustacean stomatogastric ganglion.  Also, 
they were interested in the rhythmic output of the network rather than characterizing the range of the population response heterogeneity.  
The results in \citet{marder11} are similar in spirit to what 
has been shown here in that different sets of parameters can result in similar output (also see \citet{marder06}); 
in our case, the range of excitatory firing rates can arise with different levels of heterogeneity by tuning the correlation 
between the two sources of heterogeneity.  One of the conclusions of their work is that the intrinsic and network parameters, 
or heterogeneities, must be taken as a whole and the correlation among these parameters is crucial in determining network output.  Our study compliments these assertions in a specific way, 
by determining how the correlation of two parameters alters the range of the excitatory firing rates in different regimes.

Unlike the work of \citet{hermann12,mejias12}, we do not consider how synaptic heterogeneity can induce different dynamics but rather focus on specific spiking regimes that are similar 
with and without heterogeneities.  Their work has a more detailed characterization of the dynamics, whereas our work explores the effects of two specific forms of heterogeneities away from the 
bifurcation points.  Studying how the relationship of intrinsic and network heterogeneity induces different dynamics would be interesting but is beyond the scope of this paper.  
A recent paper of \citet{ostojic14} calls the first regime that we termed noisy rhythm 'a second type of asynchronous' network but with stronger coupling.  
\citet{ostojic14} found that these two types of asynchronous networks (classic and strongly coupled) processed external stimuli differently, 
a feature that was not considered in this paper.  These two regimes (Ostojic's strongly asynchronous regime and the 
noisy rhythm here) are similar because the coupling is relatively strong and both autocorrelation functions of the population firing rates are similar.  
Consistent with illustrating that these regimes are different, we have shown how the classic asynchronous regime is different than the noisy rhythm or 
'strongly coupled asynchronous' (Ostojic) because the firing rate ranges change in distinct ways.

%Moreover, in a simple model, \citet{abbott99} showed that de-correlation might not always enhance coding.
A related study analyzes the interplay of two sources of intrinsic heterogeneity \citep{mejias14}; specifically, \citet{mejias14} studied how heterogeneity in the excitatory and inhibitory 
spiking thresholds had different effects on a coupled network (LIF with excitatory and inhibitory neurons).  They found distinct roles for heterogeneity in each type of neuron: 
excitatory heterogeneity can increase firing rate and linearizes output response, inhibitory heterogeneity can decrease network response and lead to gain control of input/output 
response (see \citet{mejias14} for further details).  Their analysis also characterized the heterogeneity-induced transitions from asynchrony to synchrony and briefly 
considered the combined effects of these two attributes.  Our work only considered excitatory heterogeneity; the effects of inhibitory neuron heterogeneity 
combined with network heterogeneity are not known and a potential future direction of research.  
Another study by \citet{hunsberger14} examined how varying both (white) noise in the voltage and heterogeneity in the 
threshold for firing led to different information (mutual information) content in spiking neuron models (LIF and Fitzhugh-Nagumo).  
They found an optimal level of heterogeneity for maximizing information content for a fixed level of noise 
(and for fixed level of heterogeneity there was an optimal level of noise, i.e., a stochastic resonance).  Their results are distinct from \citet{mejias12} and \cite{tripathy13} 
(who also found optimal information by tuning parameters) because they considered the interplay of those two sources of variability and determined that they interact nonlinearly 
(e.g., the optimal parameters are different with both components).  In our study, 
we did not systematically analyze how varying the (correlated) noise level $\sigma_E$ and heterogeneity effect network statistics.  \citet{hunsberger14} explained their simulation results by 
comparing how each component desynchronized and/or linearized the network response properties, whereas we provide an analytic explanation.  Finally, \citet{lengler13} recently 
simulated an LIF network with a large number of heterogeneous intrinsic and network parameters.  They find that heterogeneity 
can increase response time and paradoxically less variable responses (reliability), though they do not provide underlying mechanistic explanations for their results.

Our study provides a more complete understanding of 
how heterogeneities interact and result in modulation of the firing rate statistics, which may ultimately lead to a better understanding of neural coding in coupled neural networks.  
Even though the firing rate is a first order measure of the response statistics, the range of this quantity has an impact on coding.  
There have been a number of recent studies focusing on the impact of heterogeneity on neural coding.  
\citet{padmanabhan10} showed with recordings of mitral cells in mice olfactory bulb that heterogeneous cells had lower correlated activity, which is thought to 
increase information capacity of a given population.  Similarly, 
\citet{chelaru08} found diverse orientation tuning curves enhances coding with increased information via a reduction in correlated activity of a coupled LIF network.  
\citet{shamir06} proposed a theoretical 
explanation for the benefits of diversity/heterogeneity in population coding, whereby the information capacity is not limited to the correlation of activity.  
A future direction of study is how 
different forms of heterogeneities considered in this paper lead to changes in the second order statistics (correlation or co-variability), which also have implications for coding in neural systems.  
Although many of the previous studies conclude that 
heterogeneity generally leads to lower co-variability \citep{chelaru08,padmanabhan10}, better discrimination \citep{marsat10,mejias13}, and ultimately enhanced coding, 
the subtleties of how co-variability is modulated is not completely known and remains an active area of research \citep{ponce13,ruff14,mochol15}.

\begin{acknowledgements}
	We thank Gary Marsat and Brent Doiron for enlightening conversations, and Brent Doiron for providing feedback on the manuscript.  
	This work was supported by a grant from the Simons Foundation (\#355173, Cheng Ly).
\end{acknowledgements}

\section*{Appendix: Generating correlated $\vec{q}$ and $\vec{\theta}$}\label{corr_qTh}

Given two vectors $\vec{\theta}$ (intrinsic heterogeneity) and $\vec{q}$ (network heterogeneity), we can generate a new pair of vectors (of the same size) that have any desired correlation 
coefficient $\varrho\in(-1,1)$.  In this paper, we choose to keep $\vec{q}$ fixed and generate a new vector $\vec{\vartheta}$ that has the same sample mean ($\mu(\vec{\theta})$) and 
sample standard deviation ($\sigma(\vec{\theta})$) of 
$\vec{\theta}$.  Note that there are infinitely many ways to generate two such vectors if we only require that the mean and standard deviation of the new vectors be equal to the original 
statistics of the vector.  The algorithm we use is as follows.
\begin{itemize}
  \item INPUTS: ($\vec{q},\vec{\theta},\varrho$)
  \item Set $\varphi=\cos^{-1}(\varrho)$
  \item Shift input vectors so they have zero mean: 
  
  	$\vec{q}_0=\vec{q}-\mu(\vec{q})$
	
	$\vec{\theta}_0=\vec{\theta}-\mu(\vec{\theta})$.
	
  \item Calculate orthogonal complement to $\vec{q}_0$: 
  
	  $\vec{z}=\vec{\theta}_0-\frac{\vec{q}_0\cdot\vec{\theta}_0}{\|\vec{q}_0\|^2}\vec{q}_0$
  
  \item Create unit vectors out of $\vec{q}_0$ and $\vec{z}$: 
  
  	$\tilde{\vec{q}}=\vec{q}_0/\|\vec{q}_0\|$, $\tilde{\vec{z}}=\vec{z}/\|\vec{z}\|$
  
  \item Create vector with prescribed correlation and zero mean: 
  	
	$\hat{\vec{\theta}}=\cos(\varphi)\tilde{\vec{q}}+\sin(\varphi)\tilde{\vec{z}}$
	
  \item Set $\vec{\vartheta}=\frac{\sigma(\vec{\theta})}{\sigma(\hat{\vec{\theta}})}\hat{\vec{\theta}}+\mu(\vec{\theta})$
  
  \item OUTPUT: $\vec{\vartheta}$, where correlation coefficient of $\vec{\vartheta}$ and $\vec{q}$ is $\varrho$, $\mu(\vec{\vartheta})=\mu(\vec{\theta})$, and 
  $\sigma(\vec{\vartheta})=\sigma(\vec{\theta})$.
\end{itemize}

\bibliographystyle{spbasic}      % basic style, author-year citations

%\bibliography{het_spike}   % name your BibTeX data base

\end{document}